\documentstyle[12pt,epsf]{article}

\renewcommand{\thefootnote}{\fnsymbol{footnote}}

\newcommand{\ie}{{\it i.e.}}
\renewcommand{\bar}{\overline}

\newcommand{\eg}{{\it e.g.}}
\newcommand{\MS}{{\rm MS}}
\newcommand{\MSb}{\overline{\rm MS}}
\newcommand{\GeV}{{\rm GeV}}

\begin{document}
\begin{flushright}
SLAC--PUB--96--7098
\end{flushright}
\bigskip

\centerline{{\Large OPTIMAL RENORMALIZATION SCALES AND}}
\vglue.1in
\centerline{{\Large COMMENSURATE SCALE RELATIONS}
 \footnote{\baselineskip=14pt
  Work partially supported by the Department ofEnergy, contract 
  DE--AC03--76SF00515 and contract DE--FG02--93ER--40762.}}
\begin{center}
  {\large Stanley J. Brodsky}\\
 Stanford Linear Accelerator Center \\
 Stanford University, Stanford, California 94309\\
\vspace{13pt}
and \\
\vspace{13pt}
{\large Hung Jung Lu}\\ 
  Department of Physics, University of Arizona\\
  Tucson, Arizona 20742
\end{center}

\baselineskip 13pt
\vspace*{.5cm}
\begin{center}
Abstract
\end{center}
Commensurate scale relations relate observables to observables and
thus are independent of theoretical conventions, such as the choice
of intermediate renormalization scheme. The physical quantities are
related at commensurate scales which satisfy a transitivity rule
which ensures that predictions are independent of the choice of an
intermediate renormalization scheme. QCD can thus be tested in a new
and precise way by checking that the observables track both in their
relative normalization and in their commensurate scale dependence. 
For example, the radiative corrections to the Bjorken sum rule at a
given momentum transfer $Q$ can be predicted from measurements of
the $e^+ e^-$ annihilation cross section at a corresponding
commensurate energy scale $\sqrt s \propto Q,$ thus generalizing
Crewther's relation to non-conformal QCD. The coefficients that
appear in this perturbative expansion take the form of a simple
geometric series and thus have no renormalon divergent  behavior. 
We also discuss scale-fixed relations between the threshold
corrections to the heavy quark production cross section in $e^+ e^-$
annihilation and the heavy quark coupling $\alpha_V$ which is
measurable in lattice gauge theory. 

\begin{center}
Invited talk presented by SJB at the \\
International Symposium on Heavy Flavory and Electroweak Theory\\
Beijing, China --- August 16--19, 1995
\end{center}
\vglue0.6cm
\setcounter{footnote}{0}
\renewcommand{\thefootnote}{\arabic{footnote}}

\eject
\baselineskip=24pt

\vglue0.6cm
\leftline{\bf  1.~Introduction}
\vglue0.4cm

The leading power-law prediction in perturbative QCD for an
observable such as the $e^+ e^-$ annihilation cross section is
generally of the form $\Sigma^N_{n=1} r_n \alpha^n_s(\mu).$ The
predictions for a physical quantity are formally invariant under a
change of the renormalization scale $\mu$; however, since the series
is only known for a finite number of terms $N$, there is an
unavoidable dependence on the choice of $\mu.$ The choice of
renormalization scheme used to define and normalize the coupling
$\alpha_s$ is also apparently a matter of convention. In principle,
one can carry out perturbative calculations in any renormalization
scheme such as modified minimal subtraction $\alpha_{\overline {\rm
MS}}$ or by expanding in effective charges defined from any
perturbatively-calculable observable.

Thus a critical problem in making reliable predictions in quantum
chromodynamics is how to deal with the dependence of the truncated
perturbative series  on the choice of renormalization scale and
scheme. Because there is no {\it a priori} range of  allowed values
of the renormalization scale and the parameters of the
renormalization scheme, it is even difficult to quantify the
uncertainties due to the renormalization conventions. For processes
where only the leading and next-to-leading order predictions are
known,  the theoretical uncertainties from the choice of
renormalization scale and scheme are evidently much larger than the
experimental uncertainties. The problems of convention dependence
are compounded by the fact that the infinite series in PQCD is
generally divergent due to ``renormalon'' contributions growing as
$n! \alpha_s^n(\mu).$ ``Renormalons'' are singularities in the Borel
transform of asymptotic series coming from particular subset of
Feynman diagrams \cite{tHooft} which dictate the divergent behavior
of large-order expansion coefficients.

The uncertainties introduced by the conventions in the
renormalization procedure are amplified in processes involving more
than one physical scale such as jet observables and semi-inclusive
reactions. In the case of jet production at $e^+e^-$ colliders, the
the jet fractions depend both on the total center of mass energy $s$
and the jet resolution parameter $y$ (which gives an upperbound $y
s$ to the invariant mass squared of each individual jet). Kramer and
Lamper \cite{KramerLampe} have shown that different scale-setting
strategies can lead to very different behaviors for the
renormalization scale in the small $y$ region. The experimental fits
indicate that in general one must choose a scale $\mu^2 \ll s$ for
$4$-jet production rates \cite{Bethke}. In the case of QCD
predictions for exclusive processes such as the decay of heavy
hadrons to specific channels and baryon form factors at large
momentum transfer, the scale ambiguities for the underlying
quark-gluon subprocesses are even more acute since the coupling
constant $\alpha_s(\mu)$ enters at a high power.  Furthermore, since
the external momenta entering an exclusive reaction are partitioned
among the many propagators of the underlying hard-scattering
amplitude, the physical scales that control these processes are
inevitably much softer than the overall momentum transfer.

Scheme and scale ambiguities are also an obstacle for testing the
Standard Model to high precision. The situation is complicated by
the fact that computations in different sectors of the Standard
Model are carried out using different renormalization schemes.  For
example, in quantum electrodynamics, higher order radiative
corrections are computed in the traditional ``on- shell" scheme
using Pauli-Villars regularization.  The QED coupling $\alpha_{\rm
QED}$ is defined from the Coulomb scattering of heavy test charges
at zero momentum transfer. The scale $k^2$ in the running QED
coupling $\alpha_{\rm QED}(k^2)$ is then set by the virtuality of
the photon propagator in order to sum all vacuum polarization
corrections. However, in the non-Abelian sectors of the Standard
Model, higher order computations are usually carried out using the
${\overline {\rm MS}}$ dimensional regularization scheme. The
renormalization scale $\mu$ that appears in perturbative expansions
in the QCD coupling $\alpha_{\overline {\rm MS}}(\mu^2)$ is usually
treated as an arbitrary parameter.  These ambiguities and
disparities in choices of scales and schemes lead to uncertainties
in establishing the accuracy and range of validity of perturbative
QCD predictions and in testing the hypothesis of grand unification
\cite{Bagger}.

The fact that physical quantities cannot depend on theoretical
conventions does not preclude the possibility that we can choose an
optimal renormalization scale for the truncated series.  This is
analogous to gauge invariance:  physical results cannot depend on
the choice of gauge; however, there are often special gauges, such
as the radiation gauge, light-cone gauge, Landau gauge, Fried-Yennie
gauge, which allow the entire physical result to be calculated and
presented in a simple way.  Similarly, it is possible that the
renormalization scale ambiguity problem can be resolved if one can
optimize the choices of scale according to some sensible criteria.
As we shall see in the case of the generalized Crewther relation, an
appropriate choice of scales also makes the physical interpretation
more transparent.

In the BLM procedure \cite{BLM}, one first selects a renormalization
scheme. The renormalization scales are then fixed by the requirement
that all contributions to the $\beta-$function such as quark and
gluon loop contributions are re-summed into the running couplings.  
The coefficients of the perturbative series are thus identical to
the perturbative coefficients of the corresponding
conformally-invariant theory with $\beta=0.$  The BLM method has the
important advantage of ``pre-summing" the large and strongly
divergent terms in the PQCD  series which grow as $n! (\beta_0
\alpha_s)^n$, \ie, the infrared renormalons associated with
coupling- constant
renormalization \cite{tHooft,Mueller,LuOneDim,BenekeBraun}. The
renormalization scales $Q^*$ in the BLM method are physical in the
sense that they reflect the mean virtuality of the gluon
propagators \cite{BLM,LepageMackenzie,Neubert,BallBenekeBraun}.

In should be emphasized, that the BLM renormalization scales $Q^*$
and the series coefficients $r_n$ necessarily depend on the choice
of renormalization scheme since the scheme sets the units of measure
-- just as the number of inches of a given length differ from the
number of centimeters for the same length. Nevertheless,  the actual
predictions for  physical quantities are independent of the scheme
if one uses BLM scale-setting.

It is interesting to compare Pad\'e resummation predictions for
single-scale perturbative QCD series in which the initial
renormalization scale choice is taken as the characteristic scale
$\mu=Q$ as well as the BLM scale $\mu=Q^*$.  One finds \cite{BEGKS}
that the Pad\'e predictions for the summed series are in each case
independent of the initial scale choice, an indication that the
Pad\'e results are thus characteristic of the actual QCD prediction.
However, the BLM scale generally produces a faster convergence to
the complete sum than the conventional scale choice.  This can be
understood by the fact that the BLM scale choice immediately sums
into the coupling all repeated vacuum polarization insertions to all
orders, thus eliminating the large $(\beta_0\alpha_s)^n$ terms in
the series as well as the $n!$ growth characteristic of the infrared
renormalon structure of PQCD \cite{tHooft,Mueller}.

A basic principle of renormalization theory is the requirement that
relations between physical observables must be independent of
renormalization scale and scheme conventions to any fixed order of
perturbation theory \cite{StueckelbergPeterman}. In this talk, we
shall discuss high precision perturbative predictions which have no
scale or scheme ambiguities.  These predictions, called
``Commensurate Scale Relations,"\cite{CSR} are valid for any
renormalizable quantum field theory, and thus may provide a uniform
perturbative analysis of the electroweak and strong sectors of the
Standard Model.

Commensurate scale relations relate observables to observables, and
thus are independent of theoretical conventions, such as choice of
intermediate renormalization scheme. The scales of the effective
charges that appear in commensurate scale relations are fixed by the
requirement that the couplings sum all of the effects of the
non-zero $\beta$ function, as in the BLM method \cite{BLM}. The
coefficients in the perturbative expansions in the commensurate
scale relations are thus identical to those of a corresponding
conformally-invariant theory with $\beta=0.$

A helpful tool and notation for relating physical quantities is the
effective charge. Any perturbatively calculable physical quantity
can be used to define an effective
charge \cite{Grunberg,DharGupta,GuptaShirkovTarasov} by incorporating
the entire radiative correction into its definition. An important
result is that all effective charges $\alpha_A(Q)$ satisfy the
Gell-Mann-Low renormalization group equation with the same $\beta_0$
and $\beta_1;$ different schemes or effective charges only differ
through the third and higher coefficients of the $\beta$ function.
Thus, any effective charge can be used as a reference running
coupling constant in QCD to define the renormalization procedure.
More generally, each effective charge or renormalization scheme,
including $\overline{\rm MS}$, is a special case of the universal
coupling function $\alpha(Q, \beta_n)$
\cite{StueckelbergPeterman,BrodskyLu}. Peterman and St\"uckelberg
have shown \cite{StueckelbergPeterman} that all effective charges
are related to each other through a set of evolution equations in
the scheme parameters $\beta_n.$

For example, consider the entire radiative corrections to the
annihilation cross section expressed as the ``effective charge"
$\alpha_R(Q)$ where $Q=\sqrt s$:
\begin{equation}
R(Q) \equiv 3 \sum_f Q_f^2 \left[1+{\alpha_R(Q) \over \pi}
\right].
\end{equation}
Similarly, we can define the entire radiative correction to the
Bjorken sum rule as the effective charge $\alpha_{g_1}(Q)$
where $Q$
is the lepton momentum transfer:
\begin{equation}
\int_0^1
d x \left[
   g_1^{ep}(x,Q^2) - g_1^{en}(x,Q^2) \right]
   \equiv {1\over 3} \left|g_A \over g_V \right|
   \left[ 1- {\alpha_{g_1}(Q) \over \pi} \right] .
\end{equation}

The commensurate scale relations connecting the effective charges
for observables $A$ and $B$ have the form $\alpha_A(Q_A) =
\alpha_B(Q_B) \left(1 + r_{A/B} {\alpha_B\over \pi} +\cdots\right),$
where the coefficient $r_{A/B}$ is independent of the number of
flavors $n_F$ contributing to coupling constant renormalization. 
The ratio of scales $\lambda_{A/B} = Q_A/Q_B$ is unique at leading
order and guarantees that the observables $A$ and $B$ pass through
new quark thresholds at the same physical scale.  One also can show
that the commensurate scales satisfy the transitivity rule
$\lambda_{A/B} = \lambda_{A/C} \lambda_{C/B},$ which is the
renormalization group property which ensures that predictions in
PQCD are independent of the choice of an intermediate
renormalization scheme $C.$ In particular, scale-fixed predictions
can be made without reference to theoretically-constructed
renormalization schemes such as $\MSb.$ QCD can thus be tested in a
new and precise way by checking that the observables track both in
their relative normalization and in their commensurate scale
dependence.

A scale-fixed relation between any two physical observables $A$ and
$B$ can be derived by applying BLM scale-fixing to their respective
perturbative predictions in, say, the $\overline {\MS}$ scheme, and
then algebraically eliminating $\alpha_{\overline {\MS}}.$ The choice
of the BLM scale ensures that the resulting commensurate scale
relation between $A$ and $B$ is independent of the choice of the
intermediate renormalization scheme \cite{CSR}. Thus, using this
formalism, one can relate any perturbatively calculable observable,
such as the annihilation ratio $R_{e^+ e^-}$, the heavy quark
potential, and the radiative corrections to structure function sum
rules to each other without any renormalization scale or scheme
ambiguity \cite{CSR}. Commensurate scale relations can also be
applied in grand unified theories to make scale and scheme invariant
predictions which relate physical observables in different sectors
of the theory.

The scales that appear in commensurate scale relations are physical
since they reflect the mean virtuality of the gluons in the
underlying hard subprocess \cite{BLM,Neubert}. As emphasized by
Mueller \cite{Mueller}, commensurate scale relations isolate the
effect of infrared renormalons associated with the non-zero $\beta$
function. The usual factorial growth of the coefficients in
perturbation theory due to quark and gluon vacuum polarization
insertions is eliminated since such effects are resummed into the
running couplings. The perturbative series is thus much more
convergent.

In the next section we discuss an elegant example: a surprisingly
simple connection between the radiative corrections to the Bjorken
sum rule at a given momentum transfer $Q$ is predicted from
measurements of the $e^+ e^-$ annihilation cross section at a
corresponding commensurate energy scale $\sqrt s \propto
Q $\cite{CSR,BGKL}. The coefficients that appear in the perturbative
expansion are a simple geometric series, and thus have no divergent
renormalon behavior $n! \alpha_s^n(\mu)$ in the coefficients. This
relation generalizes Crewther's relation to non-conformal QCD. 
Another useful example is the connection between the moments of
structure functions and other observables \cite{BLM,Wong}.

The heavy-quark potential $V(Q^2)$ can be identified as the
two-particle-irreducible scattering amplitude of test charges; \ie \
the scattering of two infinitely-heavy quark and antiquark at
momentum transfer $t = -Q^2.$  The relation $V(Q^2) = - 4 \pi C_F
\alpha_V(Q^2)/Q^2$ with $C_F$ given by $C_F=(N_C^2-1)/2 N_C=4/3$
then defines the ``effective charge'' $\alpha_V(Q).$  This coupling
provides a physically-based alternative to the usual ${\bar {\MS}}$
scheme.

As in the corresponding case of Abelian QED, the scale $Q$ of the
coupling $\alpha_V(Q)$ is identified with the exchanged  momentum.
All vacuum polarization corrections due to fermion pairs are
incorporated in terms of the usual vacuum polarization kernels
defined in terms of physical mass thresholds.  The first two terms
$\beta_0 = 11 - {2 \over 3}n_f$ and $\beta_1 = 102 - {38 \over
3}n_f$ in the expansion of the $\beta$-function defined from the
logarithmic derivative of $\alpha_V(Q)$ are universal, \ie,
identical for all effective charges.

The scale-fixed relation between $\alpha_V$ and the conventional
$\bar {\MS}$ coupling is
\begin{equation}
\alpha_{\bar {\MS}}(Q)= \alpha_V(e^{5/6} Q)(1 + {2
\alpha_V\over\pi} +
...),
\label{EqAlphaMSAlphaV}
\end {equation}
The factor $e^{5/6} \simeq 0.4346$ is the ratio of commensurate
scales between the two schemes to this order.   It arises because of
the convention used in defining the modified minimal subtraction
scheme. The scale in the $\bar {\MS}$ scheme is thus a factor
$\sim 0.4$ smaller than the physical scale. The coefficient $2$ in
the NLO coefficient is a feature of the non Abelian couplings of
QCD; the same coefficient occurs even if the theory were conformally
invariant with $\beta_0=0.$   The commensurate scale relation
between $\alpha_V$, as defined from the heavy quark potential, and
$\alpha_{\bar {\MS}}$ provides an analytic extension of the
$\bar {\MS}$ scheme in which flavor thresholds are
automatically. taken into account at their proper respective
scales \cite{Mirabelli}.

The use of $\alpha_V$  as the expansion parameter with BLM
scale-fixing has also been found to be  valuable in lattice gauge
theory, greatly increasing the convergence of perturbative
expansions relative to those using the bare lattice coupling
\cite{LepageMackenzie}. In fact the new lattice calculations of the
$\Upsilon$- spectrum \cite{Davies} has been used to determine the
normalization of the static heavy quark potential and its effective
charge using as input a line splitting of the quarkonium spectrum:
\begin{equation}
\alpha_V^{(3)}(8.2 \GeV) = 0.196(3).
\end{equation}
where the effective number of light flavors is $n_f = 3$. The
corresponding modified minimal subtraction coupling evolved to the
$Z$ mass is given by
\begin{equation}
\alpha_{\bar{\MS}}^{(5)}(M_Z) = 0.115(2).
\end{equation}

One can also apply commensurate scale relations to the domain of
exclusive processes at large momentum transfer and exclusive weak
decays of heavy hadrons in QCD \cite{Huang}. In our new work with
Chueng-Ryong Ji and Alex Pang, we use the BLM method to fix the
renormalization scale of the QCD coupling in exclusive hadronic
amplitudes such as the pion form factor and the photon-to-pion
transition form factor at large momentum transfer.
Renormalization-scheme-independent commensurate scale relations can
then be established which connect the hard-scattering subprocess
amplitudes which control exclusive processes to other QCD
observables such as the heavy quark potential and the
electron-positron annihilation cross section. The coupling
$\alpha_V$ is particularly useful for analyzing exclusive
amplitudes. Each gluon propagator with four-momentum $k^\mu$ in the
hard-scattering quark-gluon scattering amplitude is associated with
the coupling $\alpha_V(k^2)$ \cite {BJLP,BLM}.

A direct measurement of $\alpha_V$ would in principle require the
scattering of heavy quarks. In fact, as we shall discuss in section
3, the threshold corrections to heavy quark production in $e^+ e^-$
annihilation depend directly on $\alpha_V$ at specific scales $Q^*$.
Two distinctly different scales arise as arguments of $\alpha_V$
near threshold: the relative momentum of the quarks governing the
soft gluon exchange responsible for the Coulomb potential, and a
large momentum scale approximately equal to twice the quark mass for
the corrections induced by transverse gluons. One thus can obtain a
direct determination of $\alpha_V$ the coupling in the heavy quark
potential, which can be compared with lattice gauge theory
predictions. The corresponding QED results for $\tau$ pair
production allow for a measurement of the magnetic moment of the
$\tau$ and could be tested at a future $\tau$-charm factory.

\vglue0.6cm
\leftline{\bf  2. Commensurate Scale Relations and
The Generalized Crewther}
\leftline{\hskip .2in \bf Relation in Quantum Chromodynamics}
\vglue0.4cm

In 1972 Crewther \cite{Crewther} derived a  remarkable consequence
of the operator product expansion for conformally-invariant gauge
theory.  Crewther's relation has the form
\begin{equation}
3 S = K R'
\end{equation}
where $S$ is the value of the anomaly controlling $\pi^0 \to \gamma
\gamma$ decay, $K$ is the value of the Bjorken sum rule in polarized
deep inelastic scattering, and $R'$ is the isovector part of the
annihilation cross section ratio $\sigma(e^+ e^- \to
$hadrons)/$\sigma(e^+ e^- \to \mu^+ \mu^-)$. Since $S$ is unaffected
by QCD radiative corrections \cite{Bardeen},  Crewther's relation
requires that the QCD radiative corrections to $R_{e^+ e^-}$ exactly
cancel the radiative corrections to the Bjorken sum rule order by
order in perturbation theory.

However,  Crewther's relation is only valid in the case of
conformally-invariant gauge theory, {\it i.e.} when the coupling
$\alpha_s$ is scale invariant. However, in reality the radiative
corrections to the Bjorken sum rule and the annihilation ratio are
in general functions of different physical scales. Thus Crewther's
relation cannot be tested directly in QCD unless the effects of the
nonzero $\beta$ function for the QCD running coupling are accounted
for, and the energy  scale $\sqrt s$ in the annihilation cross
section is related to the momentum transfer $Q$ in the deep
inelastic sum rules. Recently Broadhurst and Kataev
\cite{BroadhurstKataev} have explicitly calculated the radiative
corrections to the Crewther relation and have demonstrated
explicitly that the corrections are proportional to the QCD $\beta$
function.

We can use the known expressions to three loops
\cite{LarinVermaseren,GorishnyKataevLarin,SurguladzeSamuel} in
$\bar{\rm MS}$ scheme and choose the leading-order and
next-to-leading scales $Q^*$ and $Q^{**}$ to re-sum all quark and
gluon vacuum polarization corrections into the running couplings.
The values of these scales are the physical values of the energies
or momentum transfers which ensure that the radiative corrections to
each observable passes through the heavy quark thresholds at their
respective commensurate physical scales.  The final result
connecting the effective charges (see section 1) is remarkably
simple:
\begin{equation}
{\alpha_{g_1}(Q) \over \pi} = {\alpha_R(Q^*) \over \pi} -
\left( {\alpha_R(Q^{**}) \over \pi} \right)^2
+ \left( {\alpha_R(Q^{***})\over \pi} \right)^3 + \cdots .
\label{AlphaG1AlphaRAfterBLMThreeFlavors}
\end{equation}
The coefficients in the series (aside for a factor of $C_F,$ which
can be absorbed in the definition of $\alpha_s$) are actually
independent of color and are the same in Abelian, non- Abelian, and
conformal gauge theory.  The non-Abelian structure of the theory is
reflected in the scales $Q^*$ and $Q^{**}.$ Note that the a
calculational device; it simply serves as an intermediary between
observables and does not appear in the final relation
(\ref{AlphaG1AlphaRAfterBLMThreeFlavors}). This is equivalent to the
group property defined by Peterman and St\"uckelberg
\cite{StueckelbergPeterman} which ensures that predictions in PQCD
are independent of the choice of an intermediate renormalization
scheme. (The renormalization group method was developed by Gell-Mann
and Low \cite{GellMannLow} and by Bogoliubov and
Shirkov \cite{BogoliubovShirkov}).

The connection between the effective charges of observables such as
Eq. (\ref{AlphaG1AlphaRAfterBLMThreeFlavors}) is referred to as a
``commensurate scale relation" (CSR).  A fundamental test of QCD
will be to verify empirically that the observables track in both
normalization and shape as given by the CSR.  The commensurate scale
relations thus provide fundamental tests of QCD which can be made
increasingly precise and independent of the choice of
renormalization scheme or other theoretical convention. More
generally, the CSR between sets of physical observables
automatically satisfy the transitivity and symmetry properties
\cite{BrodskyLuSelfconsistency} of the scale transformations of the
renormalization ``group" as originally defined by Peterman and
St\"uckelberg \cite{StueckelbergPeterman}.  The predicted relation
between observables must be independent of the order one makes
substitutions; {\it i.e.} the algebraic path one takes to relate the
observables.

The relation between scales in the CSR is consistent with the BLM
scale-fixing procedure \cite{BLM} in which the scale is chosen such
that all terms arising from the QCD $\beta-$function are resummed
into the coupling.  Note that this also implies that the
coefficients in the perturbation CSR expansions are independent of
the number of quark flavors $f$ renormalizing the gluon propagators.
This prescription ensures that, as in quantum electrodynamics,
vacuum polarization contributions due to fermion pairs are all
incorporated into the coupling $\alpha(\mu)$ rather than the
coefficients. The coefficients in the perturbative expansion using
BLM scale-fixing are the same as those of the corresponding
conformally invariant theory with $\beta=0.$ In practice, the
conformal limit is defined by $\beta_0, \beta_1 \to 0$, and can be
reached, for instance, by adding enough spin-half and scalar quarks
as in $N=4$ supersymmetric QCD.  Since all the running coupling
effects have been absorbed into the renormalization scales, the  BLM
scale-setting method correctly reproduces the perturbation theory
coefficients of the conformally invariant theory in the $\beta \to
0$ limit.

Let us now discuss  in more detail the derivation of Eq.
(\ref{AlphaG1AlphaRAfterBLMThreeFlavors}). The perturbative series
of $\alpha_{g_1}(Q)/\pi$ using dimensional regularization and the
$\rm \bar{\MS}$ scheme with the renormalization scale fixed at
$\mu=Q$ has been computed \cite{LarinVermaseren} through three loops
in perturbation theory. The effective charge for the annihilation
cross section has also been computed
\cite{GorishnyKataevLarin,SurguladzeSamuel} to the same order in the
$\bar{\rm MS}$ scheme with the renormalization scale fixed at
$\mu=Q=\sqrt s$. The two effective charges can be related to each
other by  eliminating $\alpha_{\bar{\rm MS}}.$ The scales $Q^*$ and
$Q^{**}$ are set by resumming all dependence on $\beta=0$ and
$\beta_1$ into the effective charge. The application of the NLO BLM
formulas then leads to
\begin{eqnarray}
\frac{\alpha_{g_1}(Q)}{\pi} &=&
\frac{\alpha_R(Q^*)}{\pi}-\frac{3}{4} C_F
\left(\frac{\alpha_R(Q^{**})}{\pi}\right)^2 \nonumber\\ & &
+\left[\frac{9}{16} C_F^2-\left(\frac{11}{144}- \frac{1}{6}\zeta_3
\right)\frac{d^{abc}d^{abc}}{C_F N}\frac{\left( \sum_f Q_f\right)^2}
{\sum_f Q_f^2}\right]\left(\frac{\alpha_R(Q^{***})}{\pi}\right)^3 ,
\\ Q^* &=& Q \exp\left[\frac{7}{4}-
2\zeta_3+\left(\frac{11}{96}+\frac{7}{3}\zeta_3 -2
\zeta_3^2-\frac{\pi^2}{24}\right)\left(\frac{11}{3} C_A -\frac{2}{3}
f\right)\frac{\alpha_R(Q)}{\pi}\right],\\ Q^{**} &=& Q \exp\left[
\frac{523}{216}+\frac{28}{9} \zeta_3- \frac{20}{3} \zeta_5
+\left(-\frac{13}{54}+\frac{2}{9}
\zeta_3\right)\frac{C_A}{C_F}\right] .
\end{eqnarray}
In practice, the scale $Q^{***}$ in the above expression can be
chosen to be $Q^{**}$. Notice that aside from the light-by-light
contributions, all the $\zeta_3, \zeta_5$ and $\pi^2$ dependencies
have been absorbed into the renormalization scales $Q^*$ and
$Q^{**}$. Understandably, the $\pi^2$ term should be absorbed into
renormalization scale since it comes from the analytical
continuation of $R(Q)$ to the Euclidean region.

For the three-flavor case, where the light-by-light contribution
vanishes, the series remarkably simplifies to the CSR of Eq.
(\ref{AlphaG1AlphaRAfterBLMThreeFlavors}).  The form suggests that
for the general $SU(N)$ group the natural expansion parameter is
$\widehat\alpha= (3 C_F/4 \pi)\,\alpha$.  The use of
$\widehat\alpha$ also makes it explicit that the same formula is
valid for QCD and QED.  That is, in the limit $N_C \to 0$ the
perturbative coefficients in QCD coincide with the perturbative
coefficients of an Abelian analog of QCD.

\vspace{.5cm}
\begin{figure}[htbp]
\begin{center}
\leavevmode
{\epsfxsize=3.75truein \epsfbox{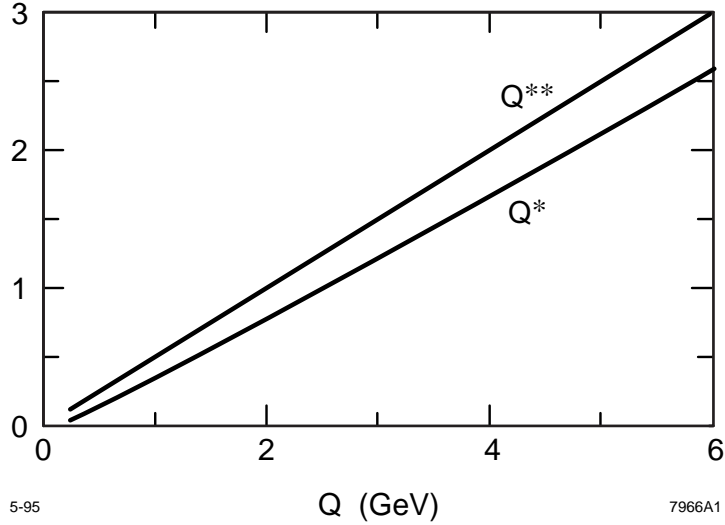}}
\end{center}
\caption[*]{The commensurate scales $Q^*$ and
$Q^{**}$ for the case of Bjorken sum rule expressed in terms of
$\alpha_R(Q)$.}
\label{fig1}
\end{figure}

In Fig. \ref{fig1} we plot the scales $Q^*$ and $Q^{**}$ as function
of $Q$ for in the range $0 \le Q \le 6$. We can see that the scales
$Q^*$ and $Q^{**}$ are of the same order as $Q$ but roughly a factor
$1/2$ to $1/3$ smaller.

\vspace{.5cm}
\begin{figure}[htbp]
\begin{center}
\leavevmode
{\epsfxsize=3.75truein \epsfbox{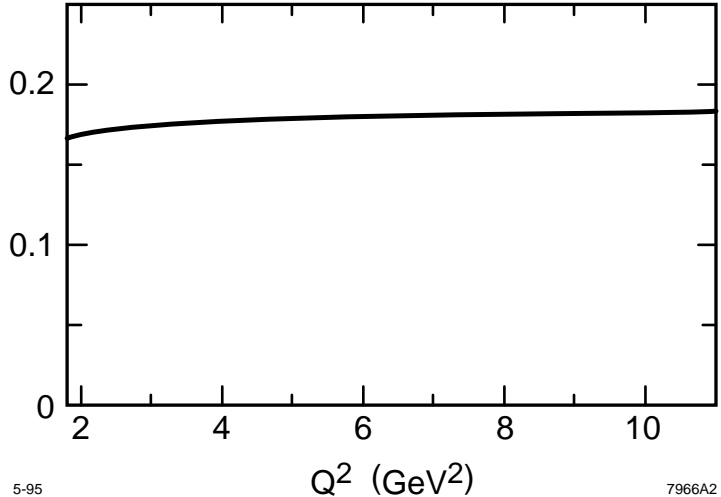}}
\end{center}
\caption[*]{Prediction of the Bjorken sum rule from
$R_{e^+e^-}$ according to the commensurate scale relation and using
Mattingly and Stevenson's result for $\alpha_R(Q)$.}
\label{fig2}
\end{figure}

In Fig. \ref{fig2} we plot the prediction for the value of the
Bjorken sum rule using as input the values of $\alpha_R(Q)$ as given
by Mattingly and Stevenson \cite{MattinglyStevenson}.  We use
$Q^{***}= Q^{**}$ here. Notice that the prediction has a very smooth
and flat behavior, even at $Q^2 \sim 2 \ {\rm GeV}^2$ since the
effective charge $\alpha_R(Q)$ as obtained by Mattingly and
Stevenson incorporates the ``freezing" of the strong coupling
constant.

Broadhurst and Kataev have recently observed a number of interesting
relations between $\alpha_R(Q)$ and $\alpha_{g_1}(Q)$ (the ``Seven
Wonders") \cite{BroadhurstKataev}.    In particular, they have shown
the factorization of the beta function in the correction to
Crewther's relation thus  establishing a non-trivial connection
between the total $e^+e^-$ annihilation cross section and the
polarized Bjorken sum rule.  The simple form of  Eq.
(\ref{AlphaG1AlphaRAfterBLMThreeFlavors}) also points to the
existence of a ``secret symmetry" between $\alpha_R(Q)$ and
$\alpha_{g_1}(Q)$ which is revealed after the application of the NLO
BLM scale setting procedure. In fact, as pointed out by Kataev and
Broadhurst \cite{BroadhurstKataev},  in the conformally invariant
limit, {\it i.e.}, for vanishing beta functions, Crewther's relation
becomes \begin{equation} (1+\widehat{\alpha}_R^{\rm eff})(1-
\widehat{\alpha}_{g_1}^{\rm eff})=1. \end {equation} Thus Eq.
(\ref{AlphaG1AlphaRAfterBLMThreeFlavors}) can be regarded as the
extension of the Crewther relation to non-conformally invariant
gauge theory.

The commensurate scale relation between $\alpha_{g_1}$ and
$\alpha_R$ given by Eq. (\ref{AlphaG1AlphaRAfterBLMThreeFlavors})
implies that the radiative corrections to the annihilation cross
section and the Bjorken (or Gross-Llewellyn Smith) sum rule cancel
at their commensurate scales.  The relations between the physical
cross sections can be written in the forms:
\begin{equation}
{R_{e^+ e^-}(s)\over 3\sum e^2_q} ~
{\int^1_0 dx  g_1^p(x,Q^2)-g_1^n(x,Q^2) \over {1\over 3}
g_A/g_V}
= 1 - \Delta \beta_0 \widehat a^3
\end{equation}
and
\begin{equation}
{R_{e^+ e^-}(s)\over 3\sum e^2_q} ~
{\int^1_0 dx  F_3^{\nu p}(x,Q^2) + F_3^{\bar\nu p}(x,Q^2)
\over 6} = 1 - \Delta \beta_0 \widehat a^3,
\label{eq10}
\end{equation}
provided that the annihilation energy in $R_{e^+ e^-}(s)$ and the
momentum transfer $Q$ appearing in the deep inelastic structure
functions are commensurate at NLO: $\sqrt s = Q^* = Q \exp [{7\over
4}- 2\zeta_3 + ({11\over 96} +{7\over 3} \zeta_3 - 2\zeta^2_3 -
{\pi^2\over 24})\beta_0 \widehat a(Q)]$. The light-by-light
correction to the CSR for the Bjorken sum rule vanishes for three
flavors.  The term $\Delta \beta_0 \widehat a^3$ with $\Delta = \ell
n\, (Q^{**}/Q^*)$ is the third-order correction arising from the
difference between $Q^{**}$ and $Q^*$; in practice this correction
is negligible: for a typical value $\widehat a = \alpha_R(Q)/ \pi =
0.14,$ $\Delta \beta_0 \widehat a^3 = 0.007.$ Thus at the magic
energy $\sqrt s = Q^*$, the radiative corrections to the Bjorken and
GLLS sum rules almost precisely cancel the radiative corrections to
the annihilation cross section.  This allows a practical test and
extension of the Crewther relation to non-conformal QCD.

As an initial test of Eq. (\ref{eq10}), we can compare the CCFR
measurement \cite{CCFR} of the Gross-Llewellyn Smith sum rule
$1-\widehat\alpha_{F_3} = {1 \over 6}\int^1_0 dx [F_3^{\nu p}(x,Q^2)
+ F_3^{\bar\nu p}(x,Q^2)] = {1\over 3} ( 2.5 \pm 0.13 )$ at
$Q^2 = 3$ GeV$^2$ and the parameterization of the annihilation data
\cite{MattinglyStevenson} $1 + \widehat\alpha_R = R_{e^+
e^-}(s)/3\sum e^2_q = 1.20.$ at the commensurate scale $\sqrt s =
Q^*= 0.38\, Q = 0.66$ GeV. The product is $(1 + \widehat\alpha_R)(1
-\widehat\alpha_{F_3})=1.00 \pm 0.04$, which is a highly nontrivial
check of the theory at very low physical scales. More recently, the
E143 \cite{E143} experiment at SLAC has reported a new value for the
Bjorken sum rule at $Q^2= 3\ $GeV$^2$: $\Gamma_1^p - \Gamma_1^n =
0.163 \pm 0.010 ({\rm stat}) \pm 0.016 ({\rm syst}).$ The Crewther
product in this case is also consistent with QCD: $(1 +
\widehat\alpha_R)(1 -\widehat\alpha_{g_1})=0.93 \pm 0.11.$

In a recent paper with Gabadadze and Kataev \cite {BGKL} we show that
it is also possible and convenient to choose one unique mean scale
$\bar Q^*$ in $\alpha_R(Q)$ so that the perturbative expansion
will also reproduce the coefficients of the geometric progression.
The possibility of using a single scale in the generalization of the
BLM prescription beyond the next-to-leading order (NLO) was first
considered by Grunberg and Kataev \cite{GrunKat}. The new single-scale
Crewther relation has the form:
\begin{equation}
\widehat{\alpha}_{g_1}(Q)=\widehat{\alpha}_R(\bar
Q^*)-
\widehat{\alpha}_R^2(\bar
Q^*)+\widehat{\alpha}_R^3(\bar Q^*)
+ \cdots,
\end{equation}

The generalized Crewther relation  provides an important test of
QCD. Since the Crewther formula written in the form of the CSR
relates one observable to another observable, the predictions are
independent of theoretical conventions, such as the choice of
renormalization scheme. It is clearly  very interesting to test
these fundamental self-consistency relations between the polarized
Bjorken sum rule or the Gross-Llewellyn Smith sum rule and the
$e^+e^-$-annihilation $R$-ratio. Present data are consistent with
the generalized Crewther relations, but measurements at higher
precision and energies will be needed to decisively test these
fundamental connections in QCD.

It is worthwhile to point out that all of the results presented here
are derived within the framework of perturbation theory in leading
twist and do not involve the nonperturbative contributions to the
Adler's function $D(Q^2)$ \cite{SVZ} and the $R$-ratio, as well as
to  the polarized Bjorken and the Gross-Llewellyn Smith sum rules 
\cite{Jaffe,BBK}.  These nonperturbative contributions are expected
to be significant at small energies and momentum  transfer. In order
to make these contributions comparatively negligible,  one should
choose relatively large values for $s$ and $Q^2$.   In order to put
the analysis of the experimental data for lower energies on more
solid ground, it will be necessary to understand whether there exist
any Crewther-type relations between non-perturbative order
$O(1/Q^4)$-corrections to the Adler's $D$-function \cite{SVZ} and
the order $O(1/Q^2)$ higher twist contributions to the
deep-inelastic sum rules \cite{Jaffe,BBK}.

The direct measurements of the polarized Bjorken sum rule (or of the
Gross-Llewellyn Smith sum rule) can be useful for the study of the
intriguing question whether the experimental data can ``sense" the
violation of the initial conformal invariance caused by the
renormalization procedure. In the language of the Crewther relation
this question can be reformulated in the following manner: what will
happen if we put the scales of $R_{e^+e^-}$ and  the corresponding
sum rules to be equal to each other? Will the experimental data
produce the conformal invariant limit, if we put $s=|Q^2|$? Recall,
that in this case, the theoretical expression for the generalized
Crewther relation will differ from the conformal invariant result
starting from the proportional to the well- known factor
$\beta(\alpha_s)/\alpha_s$ the $\alpha_s^2$-order corrections 
\cite{BK}, which presumably reflects the violation of the conformal
symmetry by the procedure of
renormalization \cite{Crewther,trace,GK}. Notice, however, that the
size of the perturbative contribution proportional to the QCD
$\beta$ - function is rather small \cite{BK}.

Commensurate scale relations such as the generalized Crewther
relation discussed here open up additional possibilities for testing
QCD.  One can compare two observables by checking that their
effective charges agree both in normalization and in their scale
dependence.  The ratio of leading-order commensurate scales
$\lambda_{A/B}$ is fixed uniquely: it ensures that both observables
$A$ and $B$ pass through heavy quark thresholds at precisely the
same physical point. The same procedure can be applied to
multi-scale problems; in general, the commensurate scales $Q^*,
Q^{**}$, etc. will depend on all of the available scales.

An important computational advantage in the commensurate scale
relations is that one only needs to compute the flavor dependence of
the higher order terms in order to specify the lower order scales in
the commensurate scale relations. We have shown \cite{CSR} that in
many cases the application of the NLO BLM formulas to relate known
physical observables in QCD leads to results with surprising
elegance and simplicity.  The commensurate scale relations for some
of the observables ($\alpha_R, \alpha_\tau, \alpha_{g_1}$ and
$\alpha_{F_3}$) are universal in the sense that the coefficients of
$\widehat \alpha_s$ are independent of color; in fact, they are the
same as those for Abelian gauge theory.  Thus much information on
the structure of the non- Abelian commensurate scale relations can
be obtained from much simpler Abelian analogs. In fact, in the
examples we have discussed here, the non-Abelian nature of gauge
theory is reflected in the $\beta$-function coefficients and the
choice of second-order scale $Q^{**}.$ The commensurate scale
relations between observables possibly can be tested at quite low
momentum transfers, even where PQCD relationships are expected to
break down.  It is possible that some of the higher twist
contributions common to the two observables are also correctly
represented by the commensurate scale relations. In contrast,
expansions of any observable in $\alpha_{\bar{\rm MS}}\,(Q)$ must
break down at low momentum transfer since $\alpha_{\bar{\rm
MS}}\,(Q)$ becomes singular at $Q=\Lambda_{\bar{\rm MS}}.$ (For
example, in the 't Hooft scheme where the higher order $\beta_n=0$
for $n=2,3,...$ , $\alpha_{\bar{\rm MS}}(Q)$ has a simple pole at
$Q=\Lambda_{\bar{\rm MS}}.$) The commensurate scale relations allow
tests of QCD in terms of finite effective charges without explicit
reference to singular schemes such as $\bar{\rm MS}.$

The coefficients in a  CSR are identical to the coefficients in a
conformal theory where explicit renormalon behavior does not appear.
It is thus reasonable to expect that the series expansions appearing
in the CSR are convergent when one relates finite observables to
each other. Thus commensurate scale relations between observables
allow tests of perturbative QCD with higher and higher precision as
the perturbative expansion grows.

\vglue0.6cm
\leftline{\bf  3.~
The Connection between the Heavy Quark Potential
and Heavy }
\leftline{\hskip.2in \bf Quark Production near Threshold}
\vglue 0.4cm

As we have noted in the introduction, the coupling $\alpha_V(Q)$ is
in many ways the natural physical coupling of QCD.   As in QED, the
scale $Q$ of this running coupling is the physical momentum
transfer.  Furthermore, $\alpha_V$  can be directly determined in
lattice gauge theory from the heavy quarkonium spectrum.

In this section, which is based on the analysis of Brodsky, Hoang,
K\"uhn, and Teubner \cite {BHKT}, we show how the heavy quark
coupling  can be measured from the angular anisotropy of heavy
quarks at threshold through the strong final-state rescattering
corrections.  We first calculate the anisotropy for QED, and then
generalize it to the non-Abelian case.   We argue that the angular
distributions of the heavy quarks will be reflected in the angular
distribution of open charm and beauty production, even in the domain
of exclusive channels.  The scale-fixed determination of the
$\bar {\MS}$ coupling is then determined using the commensurate
scale relation between $\alpha_V$ and $\alpha_{\bar {\MS}}.$

The amplitude for the creation of a massive fermion pair from a
virtual photon is characterized by the Dirac ($F_1$) and Pauli
($F_2$) form factors:
\begin{equation}
\bar u\,\Lambda_\mu\, v =
  ieQ_f\,\bar u \,\big[\, \gamma_\mu\,F_1(q^2) +
      \frac{i}{2\,m}\,\sigma_{\mu\nu}\,q^\nu\,F_2(q^2) \,
\big]\,v
\end{equation}
where  $\sigma_{\mu\nu} = \frac{i}{2}[\gamma_\mu,\gamma_\nu]$. The
photon momentum flowing into the vertex is denoted by $q$, the
fermion mass by $m$. The resulting angular distribution is
conveniently expressed in terms of the electric and magnetic form
factors $G_e$ and $G_m$ \cite{Renard}:
\begin{equation} \frac{\rm d
\,\sigma(e^+ e^- \to f\bar f)}{\rm d\,\Omega} =
 \frac{\alpha^2\,Q_f^2\,\beta}{4\,s}\bigg[
   \frac{4\,m^2}{s}\,|G_e|^2\,\sin^2\theta +
   |G_m|^2\,(1+\cos^2\theta)\, \bigg]
\end{equation}
with
\begin{equation}
G_e = F_1+\frac{s}{4\,m^2}F_2\,,\qquad
G_m = F_1+F_2\,.
\end{equation}

The anisotropy is thus given by
\begin{eqnarray}
A & = & \frac{|G_m|^2 - ( 1- \beta^2 ) |G_e|^2}
{|G_m|^2 + ( 1- \beta^2 ) |G_e|^2}
\nonumber\\
  & = & \frac{\widetilde A}{1-\widetilde A}\,,
\label{ani}
\end{eqnarray}
where
\begin{equation}
\widetilde A  =  \frac{\beta^2}{2}\,
\frac{|F_1|^2\,(1-\beta^2) - |F_2|^2}
{|F_1 + F_2|^2\,(1-\beta^2)}\,.
\label{atil}
\end{equation}

In Born approximation $A_{\rm Born} = \beta^2/(2-\beta^2)$. Notice
that for $F_2 = 0$, the anisotropy is identical to the Born
prediction, independent of $F_1$. Thus the form
\begin{equation}
\frac{\widetilde A}{\widetilde A_{\rm Born}} - 1 =
-\,2\,F_2(s)\,\bigg[\,1+{\cal
O}(\frac{\alpha}{\pi})\,\bigg]\,,\qquad
\widetilde A_{\rm Born} = \frac{\beta^2}{2}
\end{equation}
isolates $F_2(s)$. This provides a way to experimentally determine
the timelike Pauli form factor of the $\tau$ lepton. The QED
prediction is
\begin{equation} F_2(4\,m^2) = -\,\frac{\alpha}{2\,\pi} + {\cal
O}\left(\left(\frac{\alpha}{\pi}\right)^2\right)
\end{equation}
which is, up to the sign, equal to the familiar Schwinger result
$F_2(0) = \alpha/2\pi$. Away from threshold the one-loop QED
prediction is
\begin{equation} F_2(\beta) =
\frac{\alpha}{\pi}\,\bigg[\,\frac{1- \beta^2}{4\,\beta}\, \ell
n\frac{1-\beta}{1+\beta}\,\bigg]\,.
\end{equation}
This type of higher twist correction will be neglected in the
following.

In  QED,  the order $\alpha$ correction to the Dirac form factor
$\delta F_1$ in the timelike region exhibits an infrared singularity
which can be regulated using a nonvanishing photon mass $\lambda$:
\begin{equation}
\delta F_1 = \delta F_1^{\mbox{\it\scriptsize fin}} +
\delta F_1^{\mbox{\it\scriptsize IR}}\,\ell
n\frac{s}{\lambda^2}
\end{equation}
with
\begin{eqnarray}
\delta F_1^{\mbox{\it\scriptsize fin}}  & = &
\frac{\alpha\,\pi}{4\,\beta} -
\frac{3}{2}\,\frac{\alpha}{\pi} +
\frac{\alpha\,\pi}{4}\,\beta  +{\cal O}(\beta^2)\,, \\
\delta F_1^{\mbox{\it\scriptsize IR}}  & = &
-\,\frac{2}{3}\,\frac{\alpha}{\pi}\, \beta^2 +{\cal
O}(\beta^4)\,.
\end{eqnarray}
The leading term of $F_1^{\mbox{\it\scriptsize fin}}$ is
proportional $\pi\alpha/\beta$ and exhibits the familiar Coulomb
singularity. Also the constant term and the term linear in $\beta$
are infrared finite. The infrared singular part of $F_1$ is strongly
suppressed at threshold $\propto \beta^2$, giving rise to a
$\beta^3$ contribution to the rate. The correction to the Pauli form
factor $\delta F_2$ is infrared finite and approaches a constant
value at threshold:
\begin{equation} \delta F_2 =
-\,\frac{1}{2}\,\frac{\alpha}{\pi}  +{\cal O}(\beta^2)\,.
\end{equation}
Real radiation, in contrast, vanishes as $\beta^3$ in the threshold
region, where two powers of $\beta$ result from the square of the
dipole matrix element, and one power of $\beta$ comes from phase
space. It  exhibits the same logarithmic dependence on the infrared
cutoff as the $F_1$ form factor and the same leading $\beta$
dependence as the infrared singular part of the virtual correction.
As a  consequence of the strong suppression $\propto \beta^3$ it can
be neglected in the threshold region, together with the
corresponding infrared divergent part of the form factor. The
angular distribution and, similarly, the correction to the total
cross section in the threshold region are therefore determined by
the the infrared finite parts of the form factors. To order $\alpha$
one thus finds for the coefficient describing the angular dependent
piece
\begin{equation} A = A_{\rm Born}\, \left( \, 1 + \frac{2}{2 -
\beta^2}\, \frac{\alpha}{\pi} \,\right) \,.
\end{equation}
As we shall show there are interesting modifications of the
anisotropy due to the running of the QCD coupling, and the
dependence of the renormalization scale on $\sqrt{s}$ and $|\vec
q\,|$ will be crucial.

The ${\cal O}(\alpha^2)$-QED corrections to the form factors,
induced by light fermion loops, have been calculated analytically
\cite{HKT1}. In the threshold region one obtains
\begin{eqnarray}
F_1 & = &
1 + \frac{\alpha\,\pi}{4\,\beta}\,
\left[\, 1 + \bigg(\frac{\alpha}{\pi}\bigg)\,
\sum_{i = 1}^{n_f}\frac{1}{3}\,\left( \ell
n\frac{s\,{{\beta}^2}}{m_i^2
} - \frac{8}{3} \right) \,  \right] \,  \nonumber\,\\
 & & \mbox{}\,\,\, -
  \frac{3}{2}\,\frac{\alpha}{\pi}\,
   \left[\, 1 + \bigg(\frac{\alpha}{\pi}\bigg)\,
       \sum_{i = 1}^{n_f}
         \frac{1}{3}\,\left( \ell n\frac{s}{4\,m_i^2} -
            \frac{1}{2} \right)  \,  \right] \,,\\
F_2 & = &
\quad \ \ \frac{\alpha\,\pi}{4\,\beta}\,\left[ \,
   \bigg(\frac{\alpha}{\pi}\bigg)\,\frac{n_f}{3}\,\right]
   \nonumber\,\\
 & & \mbox{}\,\,\, -
  \frac{1}{2}\,\frac{\alpha}{\pi}\,
   \left[\, 1 + \bigg(\frac{\alpha}{\pi}\bigg)\,
       \sum_{i = 1}^{n_f}
         \frac{1}{3}\,\left( \ell n\frac{s}{4\,m_i^2} -
            \frac{13}{6} \right)  \, \right] \,.
\end{eqnarray}
The calculation has been performed in the limit where  the mass of
the light virtual fermion $m_f$ is far smaller than $m$, a situation
appropriate for the subsequent translation to QCD. The factor $n_f$
is introduced to allow for several light fermions and, in our case,
to single out the fermion-induced terms. These formulae provide the
first step on the way to a full two-loop calculation in order
$\alpha^2$. As we shall see, the results require two conceptually
different scales in the argument of the running coupling, a scale of
order $s$ from the hard virtual correction from transverse photons
and a soft scale of order $s\,\beta^2$ from the Coulomb
rescattering. Supplemented by the BLM prescription they even
determine the dominant two-loop gluon-induced terms in QCD.

The linear combination appearing in the denominator of $\widetilde
A$ in Eq.~(\ref{atil}) is thus given by
\begin{eqnarray}
F_1+F_2 & = &
1 + \frac{\alpha\,\pi}{4\,\beta}\,
   \left[\, 1 + \bigg(\frac{\alpha}{\pi}\bigg)\,
      \sum_{i = 1}^{n_f}
         \frac{1}{3}\,\left( \ell n\frac{s\,{{\beta}^2}}{m_i^2
              } - \frac{5}{3} \right)  \,  \right] \,
   \nonumber\,\\
 & & \mbox{}\,\,\, -
  2\,\frac{\alpha}{\pi}\,
   \left[\, 1 + \bigg(\frac{\alpha}{\pi}\bigg)\,
      \sum_{i = 1}^{n_f}
         \frac{1}{3}\,\left( \ell n\frac{s}{4\,m_i^2} -
            \frac{11}{12} \right) \,  \right] \,.
\end{eqnarray}
The $n_f$ terms arise from the vacuum polarization
insertions and
thus can be resummed into the QED running coupling:
\begin{equation}
\alpha(Q^2) = \alpha\,\left[\,
 1 + \bigg(\frac{\alpha}{\pi}\bigg)
      \sum_{i = 1}^{n_f}
         \frac{1}{3}\,\left( \ell n\frac{Q^2}{m_i^2} -
            \frac{5}{3} \right) \,\right]\,.
\end{equation}
The constant 5/3 is the usual term in the Serber-Uehling vacuum
polarization $\Pi(Q^2)$ at large $Q^2$. This corresponds to the
usual QED scheme where $V(Q^2)=-\,4\pi\,\alpha(Q^2)/Q^2$ is the QED
potential for the scattering of heavy test charges. One thus obtains
\begin{eqnarray}
F_1 + F_2 & = &
1 + \frac{\alpha(s\,\beta^2)\,\pi}{4\,\beta} -
2\,\frac{\alpha(s\,e^{3/4}/4)}{\pi} \nonumber \\
& \cong & \mbox{} \left(1-
2\,\frac{\alpha(s\,e^{3/4}/4)}{\pi}\right)\,
\left( 1 + \frac{\alpha(s\,\beta^2)\,\pi}{4\,\beta} \right)\,.
\end{eqnarray}
Two distinctly different correction factors arise. The first
originates from hard transverse photon exchange, with the scale set
by the short distance process; the second from the instantaneous
Coulomb potential. It is remarkable and non-trivial that the
non-logarithmic terms in the $\pi\alpha/\beta$ corrections are
absorbed if the relative momentum is adopted as the scale for the
coupling. Up to two loops the running coupling governing the Coulomb
singularity is thus identical to the running coupling in the
potential. This will provide an important guide for the application
of these results to QCD.

The proper resummation of the $1/\beta$ terms based on Sommerfeld's
rescattering formula then leads to
\begin{equation}
|F_1+F_2|^2 =
\left(1- 4\,\frac{\alpha(m^2\,e^{3/4})}{\pi}\right)\,
\frac{x}{1-e^{-x}}
\end{equation}
with
\begin{equation}
x = \frac{\alpha(4\,m^2\,\beta^2)\,\pi}{\beta}\,.
\end{equation}
In a similar way one finds for the relevant combination in the
numerator of (\ref{atil})
\begin{equation}
|F_1|^2-|F_2|^2 \cong
\left(1- 3\,\frac{\alpha(m^2\,e^{7/6})}{\pi}\right)\,
\frac{x^\prime}{1-e^{-x^\prime}}
\end{equation}
with
\begin{equation}
x^\prime = \frac{\alpha(4\,m^2\,\beta^2/e)\,\pi}{\beta}\,.
\end{equation}
The $|F_2|^2$ term in the numerator can actually be ignored in the
present approximation. The scales of the effective coupling differ
in the numerator and denominator of (\ref{atil}). In particular,
in the factor arising from Coulomb exchange the scale is
significantly smaller in the numerator than in the denominator. This
behavior is consistent with  qualitative considerations based on the
relative distances relevant for $S$- versus $P$-waves in the Coulomb
part. In the factor arising from hard photon exchange the scales are
quite comparable, with a slightly larger value in the numerator.

One thus arrives at the prediction in QED for the anisotropy which
involves four scales:
\begin{equation}
A = \frac{\widetilde A}{1-\widetilde A}\,,\qquad
\widetilde A = \frac{\beta^2}{2}\,
 \frac{\left(1-3\,
\frac{\alpha(m^2 e^{7/6})}{\pi}\right)}
      {\left(1-4\,
      \frac{\alpha(m^2 e^{3/4})}{\pi}\right)}\,
 \frac{1-e^{-x}}{1-e^{-x^\prime}}\,
 \frac{\alpha(4\,m^2\,\beta^2/e)}{\alpha(4\,m^2\,\beta^2)}\,.
\end{equation}
\begin{figure}
\begin{center}
\leavevmode
\epsfxsize=12cm
\epsffile[120 300 450 530]{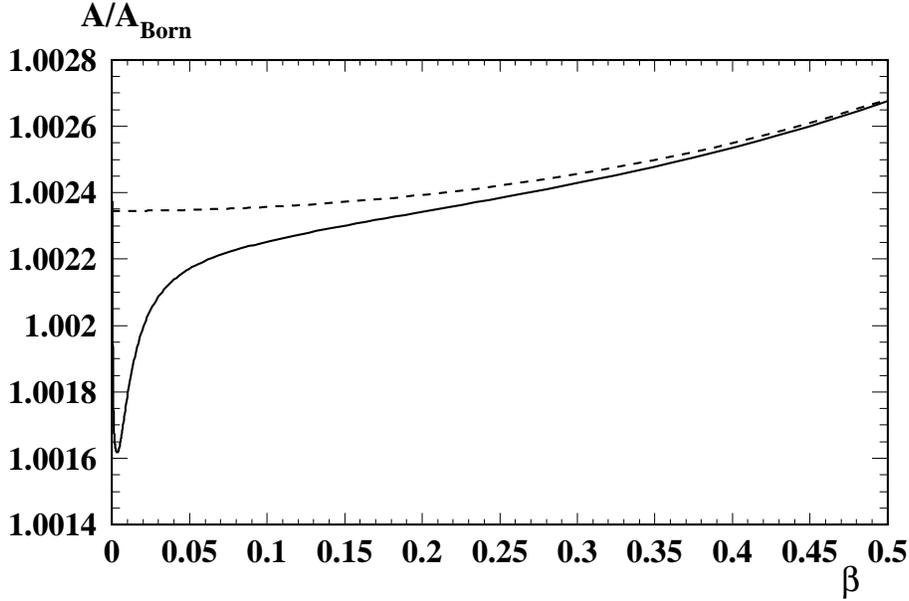}
\vskip -3mm
\caption[]{{Ratio between the anisotropy $A$ and the
Born prediction $A_{\rm Born}$ as function of $\beta$ for the
process $e^+ e^- \to \tau^+ \tau^-$. Dashed curve: constant
$\alpha$; solid curve: including the running of $\alpha$.}}
\label{fig1a}
\end{center}
\end{figure}
\noindent
To display the effects more clearly, the ratio of the anisotropy to
the Born prediction $A/A_{\rm Born}$ is shown in Fig.~\ref{fig1a}
for the case of $\tau$ pair production. The dashed curve gives the
prediction for constant $\alpha_{\rm QED}$; the solid curve shows
the effect of the lepton vacuum polarization $\Pi(Q^2)$ in the QED
running coupling. The vacuum polarization affects the anisotropy for
small $\beta$ because two different scales appear in the $S$- and
$P$-wave Coulomb rescattering corrections. Away from threshold $A$
essentially measures the anomalous magnetic moment.

The QED coupling $\alpha(Q^2)$ translates into the the QCD coupling
$\alpha_{\mbox{\it\scriptsize V}}(Q^2)$, defined as the effective
charge in the potential
\begin{equation} V(Q^2) = -\,\frac{4}{3}\,\frac{4\,\pi\,
\alpha_{\mbox{\it\scriptsize V}}(Q^2)}{Q^2}
\end{equation}
for the scattering of two heavy quarks in a color-singlet state. In
the BLM procedure all terms arising from the non-zero beta- function
are resummed into $\alpha_{\mbox{\it\scriptsize V}}(Q^2)$. For
example, all $n_f$-dependent coefficients vanish in the
$\pi\alpha/\beta$ terms if the scale of the relative momentum is
adopted. This is, in fact, a result expected on general grounds:
threshold physics is governed by the nonrelativistic instantaneous
potential. Below threshold, the potential leads to bound states,
above threshold it affects the cross section through final state
interactions. It is, therefore, natural to take for the QCD case the
coupling governing the QCD potential at the momentum scale involved
in the rescattering.

In a similar way, BLM scale-fixing is adopted for the correction
from hard gluon exchange. In the radiative correction, there still
remain ${\cal O}(\alpha_s^2)$ terms, identical to the radiative
corrections for the theory with a fixed coupling constant. With the
same scheme convention for the coupling as above, one arrives at
\begin{equation} \widetilde A = \frac{\beta^2}{2}\,
 \frac{\left(1-4\,
\frac{\alpha_{\mbox{\it\tiny V}}(m^2 e^{7/6})}{\pi}\right)}
      {\left(1-\frac{16}{3}\,
      \frac{\alpha_{\mbox{\it\tiny V}}(m^2
e^{3/4})}{\pi}\right)}\,
 \frac{1-e^{-x_s}}{1-e^{-x_s^\prime}}\,
 \frac{\alpha_{\mbox{\it\tiny V}}(4\,m^2\,\beta^2/e)}
      {\alpha_{\mbox{\it\tiny V}}(4\,m^2\,\beta^2)}
\end{equation}
where
\begin{equation}
x_s = \frac{4\,\pi}{3}\,
 \frac{\alpha_{\mbox{\it\scriptsize
V}}(4\,m^2\beta^2)}{\beta}\,,
 \qquad
x_s^\prime =
\frac{4\,\pi}{3}\,\frac{\alpha_{\mbox{\it\scriptsize V}}
 (4\,m^2\beta^2/e)}{\beta}\,.
\end{equation}
\begin{figure}
\begin{center}
\leavevmode
\epsfxsize=12cm
\epsffile[120 300 450 530]{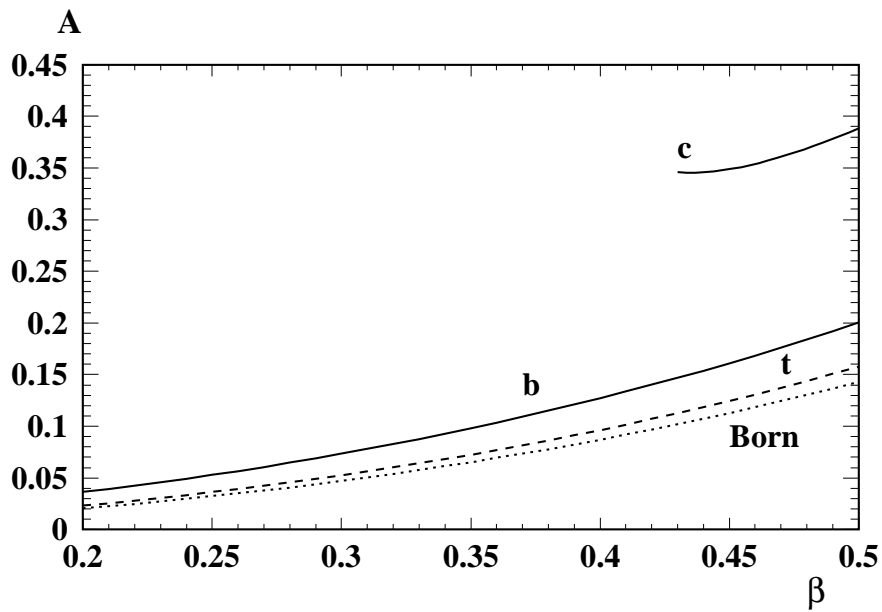}
\vskip -3mm
\caption[]{{Anisotropy for charmed, bottom and top
quark production as a function of $\beta$. Also shown is the Born
prediction. We have assumed the effective quark masses $m_c = 1.7$
GeV, $m_b = 5$ GeV and $m_t = 175$ GeV.}}
\label{fig2a}
\end{center}
\end{figure}
The anisotropy $A$ is plotted in Fig.~\ref{fig2a} versus the
velocity $\beta$ in the range  $0.2 <\beta < 0.5$ for charmed,
bottom, and top quarks. For comparison, the tree level prediction is
also shown. For charmed quarks, only $\beta$ values above $0.4$ are
admitted in order to allow for the simultaneous production of $D
\bar D$ and $D^*\bar D$. The charm prediction is particularly
sensitive to the QCD parameters, since very low scales are
accessible. Measurements of the anisotropy for $e^+ e^- \to c \bar
c$ thus have the potential of determining
$\alpha_{\mbox{\it\scriptsize V}}$ in the regime where perturbation
theory begins to fail.
\begin{figure}
\begin{center}
\leavevmode
\epsfxsize=12cm
\epsffile[120 300 450 530]{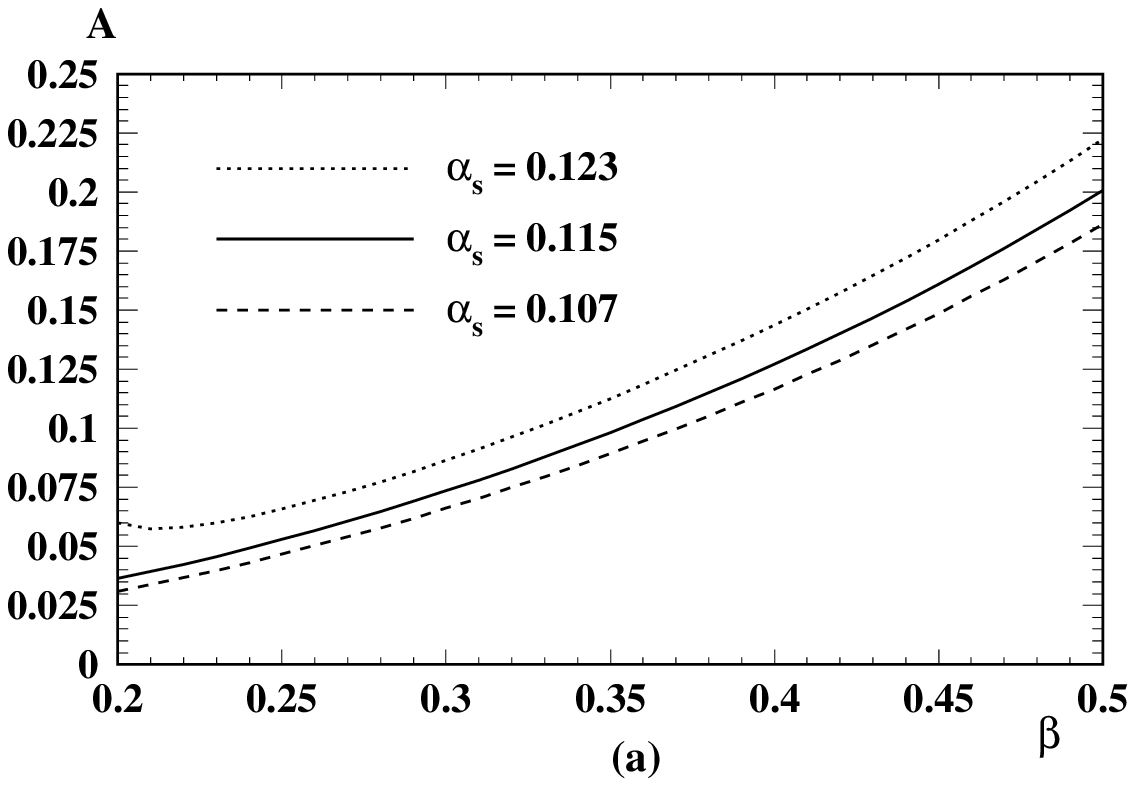}\\
\leavevmode
\epsfxsize=12cm
\epsffile[120 300 450 530]{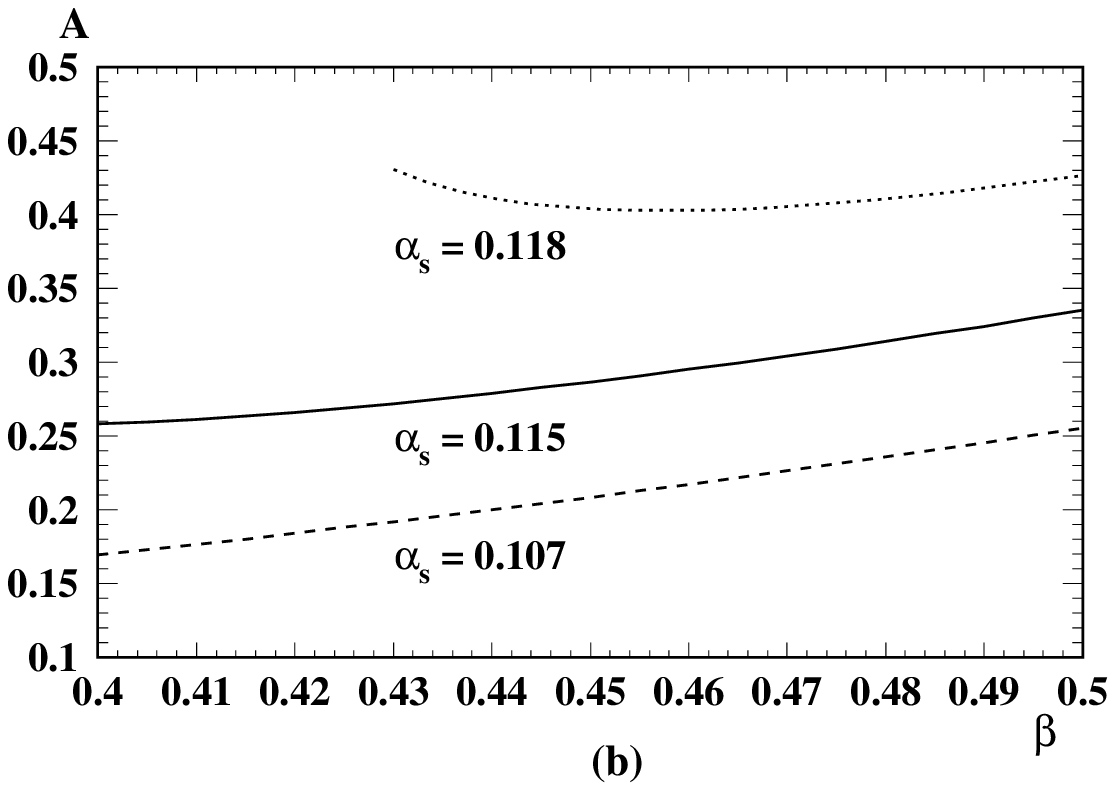}
\vskip -3mm
\caption[]{{Sensitivity of the anisotropy $A$ for
(a) $e^+ e^- \to b\,\bar b$ and (b) $e^+ e^- \to c\,\bar c$ to
changes in $\alpha_{\bar{\MS}}(M^2_Z)$.}}
\label{fig3}
\end{center}
\end{figure}

The curves are based on an input value
$\alpha_{\bar{\MS}}^{(n_f=5)}(M_Z^2) = 0.115$.  We use the
two-loop beta-function to evolve $\alpha_{\bar{\MS}}$ to lower
momenta and then used Eq.~(\ref{EqAlphaMSAlphaV}) to calculate
$\alpha_{\mbox{\it\scriptsize V}}(Q^2)$. To investigate the
sensitivity of the predictions for bottom quarks, the input value
for $\alpha_{\bar{\MS}}(M_Z^2)$ has been varied by $\pm 0.008$
from the central value of 0.115. As demonstrated in Fig.~\ref{fig3}a
the variation of the anisotropy parameter amounts to about 10\%, and
could therefore be accessible experimentally. The charm predictions
(see Fig.~\ref{fig3}b) are even more sensitive.

The anisotropy $A(\beta^2)$ in the center-of-mass angular
distribution ${\rm d}\sigma(e^+ e^- \to Q\bar Q)/{\rm d}\Omega
\propto 1+A\,\cos^2\theta$ of heavy quarks produced near threshold
is sensitive to the QCD coupling $\alpha_{\mbox{\it\scriptsize
V}}(Q^2)$ at specific scales determined by the quark relative
momentum $p_{\rm cm} = \sqrt{s}\,\beta$. The coupling
$\alpha_{\mbox{\it\scriptsize V}}(Q^2)$ is the physical effective
charge defined through heavy quark scattering. An important
consequence of heavy quark kinematics is that the production angle
of a heavy hadron follows the direction of the parent heavy quark.
This applies not only at Born approximation, but also after QCD
corrections have been applied.  The predictions thus provide a
connection between two types of observables, the heavy quark
potential and the angular distribution of heavy hadrons, independent
of theoretical scale and scheme conventions.

An important feature of this analysis is the use of BLM scale-%
fixing, in which all higher-order corrections associated with the
beta-function are resummed into the scale of the coupling. The
resulting scale for $\alpha_{\mbox{\it\scriptsize V}}(Q^2)$
corresponds to the mean gluon virtuality. In the case of the soft
rescattering corrections to the $S$-wave, the BLM scale is
$s\,\beta^2 = p^2_{\rm cm}$. One thus has sensitivity to the running
coupling over a range of momentum transfers within the same
experiment. The anisotropy measurement thus can provide a check on
other determinations of $\alpha_{\mbox{\it\scriptsize V}}(Q^2)$, \eg
\ from heavy quark lattice gauge theory, or from the conversion of
$\alpha_{\bar{\MS}}$ determinations to $\alpha_{\mbox{\it\scriptsize
V}}.$

Our analysis also shows that the running coupling appears within the
cross section with several different scales. This is particularly
apparent at low $\beta$ where the physical origin of the ${\cal
O}(\alpha_s)$ corrections can be traced to gluons with different
polarization and virtuality.

In principle, the anisotropy of $\tau$ pairs produced in $e^+e^-
\rightarrow \tau^+\tau^-$ could be used to measure the Pauli form
factor $F_2(s)$ near threshold $s\
{\scriptscriptstyle\stackrel{\scriptscriptstyle >}{ \sim}}\
4m^2_\tau$. A highly precise measurement of the anisotropy thus
could provide a measurement of a fundamental parameter of the $\tau$
lepton: its timelike anomalous magnetic moment.

\begin{sloppy}
\begin{raggedright}
\def\app#1#2#3{{\it Act. Phys. Pol. }{\bf B #1} (#2) #3}
\def\apa#1#2#3{{\it Act. Phys. Austr.}{\bf #1} (#2) #3}
\def\lhc{Proc. LHC Workshop, CERN 90-10}
\def\npb#1#2#3{{\it Nucl. Phys. }{\bf B #1} (#2) #3}
\def\plb#1#2#3{{\it Phys. Lett. }{\bf B #1} (#2) #3}
\def\prd#1#2#3{{\it Phys. Rev. }{\bf D #1} (#2) #3}
\def\pR#1#2#3{{\it Phys. Rev. }{\bf #1} (#2) #3}
\def\prl#1#2#3{{\it Phys. Rev. Lett. }{\bf #1} (#2) #3}
\def\prc#1#2#3{{\it Phys. Reports }{\bf #1} (#2) #3}
\def\cpc#1#2#3{{\it Comp. Phys. Commun. }{\bf #1} (#2) #3}
\def\nim#1#2#3{{\it Nucl. Inst. Meth. }{\bf #1} (#2) #3}
\def\pr#1#2#3{{\it Phys. Reports }{\bf #1} (#2) #3}
\def\sovnp#1#2#3{{\it Sov. J. Nucl. Phys. }{\bf #1} (#2) #3}
\def\jl#1#2#3{{\it JETP Lett. }{\bf #1} (#2) #3}
\def\jet#1#2#3{{\it JETP Lett. }{\bf #1} (#2) #3}
\def\zpc#1#2#3{{\it Z. Phys. }{\bf C #1} (#2) #3}
\def\ptp#1#2#3{{\it Prog.~Theor.~Phys.~}{\bf #1} (#2) #3}
\def\nca#1#2#3{{\it Nouvo~Cim.~}{\bf #1A} (#2) #3}
\bigskip
{\bf\noindent  Acknowledgments \hfil}
\vglue 0.4cm
The work on the generalized Crewther relation that is reported in
Section 2 has greatly benefitted from a collaboration with A. Kataev
and G. Gabadadze.   The results in section 3 is based on an analysis
of Brodsky, Hoang, K\"uhn, and Teubner.  We also thank M. Beneke, V.
Braun, L. Dixon, M. Gill, J. Hiller, P. Huet, C.-R. Ji, G.  P.
Lepage, G. Mirabelli, A.  Mueller, D. M\"uller, Alex Pang,
O.~Puzyrko,  and W.-K.  Wong, for helpful discussions. We wish to
thank Tao Huang, the Institute for High Energy Physics and the
organizers of the International Symposium on Heavy Flavor and
Electroweak theory for their hospitality in Beijing. This work is
supported in part by the Department of Energy, contract DE--%
AC03--76SF00515 and contract DE--FG03--93ER--40792.

\end{raggedright}
\end{sloppy}

\end{document}